# Aluminum oxide nucleation in the early stages of atomic layer deposition on epitaxial graphene

E. Schilirò[1], R. Lo Nigro[1], S. E. Panasci[2,1], F. M. Gelardi[3], S. Agnello[3,1], R. Yakimova[4], F. Roccaforte[1], F. Giannazzo[1,*]

[1] *Consiglio Nazionale delle Ricerche – Istituto per la Microelettronica e Microsistemi (CNR-IMM), Strada VIII, 5 95121, Catania, Italy*

[2] *Department of Physics and Astronomy, University of Catania, Via Santa Sofia 64, 95123 Catania, Italy*

[3] *Department of Physics and Chemistry Emilio Segrè, University of Palermo, Via Archirafi 36, 90143 Palermo, Italy*

[4] *Department of Physics, Chemistry and Biology, Linköping University, Linköping SE-58183, Sweden*

**\*** Corresponding author. E-mail: filippo.giannazzo@imm.cnr.it  (F. Giannazzo)

**Abstract**

In this work, the nucleation and growth mechanism of aluminum oxide ($Al_2O_3$) in the early stages of the direct atomic layer deposition (ALD) on monolayer epitaxial graphene (EG) on silicon carbide (4H-SiC) has been investigated by atomic force microscopy (AFM) and Raman spectroscopy. Contrary to what is typically observed for other types of graphene, a large and uniform density of nucleation sites was observed in the case of EG and ascribed to the presence of the buffer layer at EG/SiC interface. The deposition process was characterized by $Al_2O_3$ island growth in the very early stages, followed by the formation of a continuous $Al_2O_3$ film (~2.4 nm thick) after only 40 ALD cycles due to the islands coalescence, and subsequent layer-by-layer growth. Raman spectroscopy analyses showed low impact of the ALD process on the defects density and doping of EG. The EG strain was also almost unaffected by the deposition in the regime of island growth and coalescence, whereas a significant increase was observed after the formation of a compact Al$_2$O$_3$ film. The obtained results can have important implications for device applications of epitaxial graphene requiring the integration of ultra-thin high-k insulators.

**Keywords:** Atomic layer deposition, epitaxial graphene, atomic force microscopy, Raman spectroscopy; nucleation.

1. Introduction

Due to its excellent structural, optical and electronic properties, graphene has been the object of an ever-increasing interest in the last 15 years both for fundamental studies and for a wide range of applications, including ultra-high frequency transistors [1,2], environmental or bio sensors [3] and quantum metrology standards [4]. The integration of thin insulating layers on the graphene surface represents a key step for the fabrication of graphene-based electronic devices. In particular, high-k dielectrics, such as aluminum oxide (Al$_2$O$_3$) or hafnium oxide (HfO$_2$) have been widely explored as gate insulators for graphene field effect transistors [1,2,5] and, more recently, as tunneling barriers of vertical hot electron transistors with a graphene base [6,7,8]. Furthermore, they can work as protective/encapsulation films or as functionalization layers for graphene-based sensors [9,10].

Deposition methods that are able to provide high quality ultra-thin insulators with conformal and uniform coating on large area are required for new-generation ultra-scaled devices. In this context, the atomic layer deposition (ALD) is the most appropriate technique to obtain films of high-k insulators with these features. In fact, due to its peculiar deposition mechanism consisting of sequential and self-limited reactions between the chemical precursors and the substrate surface, the ALD allows, in principle, a layer-by-layer growth, resulting in a linear increase of the film thickness with the number of deposition cycles [11,12,13,14]. However, significant deviations from this ideal behavior, especially in the early stages of the deposition, can be observed in some experimental cases

[15], depending on the used ALD precursors and on the specific substrate. As an example, a sub-monolayer growth per cycle can be due to steric effects from the precursors' ligands that block active sites, or to competing chemisorption pathways [15]. In the case of substrates with low density of nucleation sites, the 3D growth of islands or nanoparticles is typically observed instead of a 2D layer-by-layer growth. The 3D growth may occur either by attachment of material to the originally formed island, or by diffusion phenomena of the deposited material (adatoms and islands) [15]. In these cases, it is necessary to protract the ALD process over a large number of deposition cycles to observe the islands coalescence and the full layer formation [11,16].

Due to its planar $sp^2$ structure without out-of-plane bonds, graphene is an example of a surface highly unfavorable to ALD nucleation. In most of the cases, the ALD growth on graphene is very inhomogeneous, and is limited to graphene defects and edges regions [17], or in correspondence of graphene domain boundaries and wrinkles [18,19]. For these reasons, chemical pre-functionalization treatments using plasma, reactive gases [20,21] or graphene coating with different kinds of seeding-layers [22,23,24,25,26] are commonly employed to enable the ALD nucleation. Although such procedures are effective to get uniform ALD growth on graphene, they can affect, to some extent, its structural and electronic properties. Furthermore, the interfacial seed layer can be responsible for charge trapping phenomena at the dielectric/graphene interface and for an increased equivalent oxide thickness (EOT) of the seed layer/insulator stack [26]. For this reason, the direct (i.e. pre-functionalization - and seed-layer-free) ALD of insulators on graphene remains highly desirable. To date, the possibility of growing uniform high-k dielectric films directly on graphene surface by thermal ALD has been demonstrated in two specific cases, i.e. for CVD grown monolayer (1L) graphene residing on the native metal surface [27] and in the case of 1L epitaxial graphene (EG) on SiC(0001) [28]. The enhanced nucleation on these high quality (i.e. low defects density) graphene materials was ascribed, in part, to the peculiar properties of the graphene interface with the native substrates and to the transparency of graphene to electric fields generated by interfacial charges or dipoles beneath graphene [29].

In particular, due to its peculiar formation mechanisms, EG obtained by thermal decomposition of SiC(0001) [30,31,32] is characterized by the presence of an interfacial carbon layer (buffer layer) having partial sp$^3$ hybridization with the Si face. This unique interface structure is responsible for a compressively strained EG. Moreover, the electrostatic interaction with the positively charged dangling bonds at the buffer layer/SiC interface are the cause of a high n-type doping (in the order of $10^{13}$ cm$^{-2}$) of the overlying EG. Uniform, conformal and compact $Al_2O_3$ layers with a thickness ~12 nm and a breakdown field of 8 MV/cm have been recently obtained directly onto 1L EG on SiC (0001) by a standard ALD at 250 °C using TMA and water as the Al precursor and co-reactant, respectively [28]. An important role for the uniform $Al_2O_3$ growth onto 1L EG was ascribed to the electrostatic doping induced by the buffer layer/SiC dangling bonds. In particular, according to ab-initio calculations, the n-type doping improves the adsorption energy of the water molecules, namely of the ALD co-reactant, providing a larger number of reactive sites for $Al_2O_3$ formation during the subsequent pulses of the Al precursor. This electrostatic effect is fairly homogeneous on the entire 1L EG surface, whereas it weakens with increasing the number of graphene layers due to the higher shielding effect from a multilayer system, resulting in a less homogenous ALD growth on the thicker EG regions [28].

Clearly, the direct ALD growth of $Al_2O_3$ on 1L EG can have important implications, especially for advanced electronic applications requiring the integration of ultra-thin insulating layers on top of graphene. However, the deposition of relatively thick (> 10 nm) $Al_2O_3$ films by direct ALD on EG has been considered so far [33], and a clear understanding of the nucleation and growth mechanisms of the $Al_2O_3$ films on the EG/SiC system in the early stages of the ALD process is still lacking.

In this work, the morphology evolution of $Al_2O_3$ deposited by ALD on 1L EG samples has been investigated over a wide range of deposition cycles (from 10 to 190). $Al_2O_3$ island growth from a very large density of nucleation sites was observed in the early stages of the ALD process, followed by the formation of a continuous $Al_2O_3$ film (~2.4 nm thick) after only 40 cycles by coalescence of the islands. Raman spectroscopy analyses showed low impact of the direct ALD growth on the doping of EG. The native compressive strain of EG was almost unaffected by the deposition in the regime of island growth and coalescence, whereas a significant increase was observed after the formation of a compact $Al_2O_3$ film.

## 2. Experimental details

The EG samples used in this study were obtained by thermal decomposition of "nominally" on-axis 4H-SiC in a sublimation reactor at a temperature of 2000 °C and at a pressure of 900 mbar in Ar ambient. These growth conditions result in monolayer (1L) EG coverage on a very large percentage (>98%) per SiC surface area [28].

The $Al_2O_3$ films were deposited by thermal ALD processes in a PE-ALD LL SENTECH Instruments GmbH reactor, using TMA and $H_2O$ as Al and O sources, respectively. Both TMA and $H_2O$ precursors are delivered from the cylinders to the reactor chamber by $N_2$ carrier gas with a flow rate of 40 sccm. The ALD cycle consists of TMA and $H_2O$ pulses of 0.06 s and 1 s, respectively, each of two followed by 2s of purging pulse by $N_2$ to remove unreacted precursors. All ALD growths have been carried out using a deposition temperature of 250 °C and a chamber pressure of 20 Pa.

Morphological analyses were carried out by tapping mode AFM using a DI3100 equipment by Bruker with Nanoscope V electronics. Sharp silicon tips with a curvature radius of 5 nm were used for these investigations.

Raman spectroscopy measurements were conducted using a Bruker Senterra micro-Raman system equipped with a confocal microscope and a 532 nm excitation laser. A nominal laser power of 5 mW was applied to avoid damaging of EG. A pseudo-confocal aperture of 50×1000 μm was used for these analyses and the spectra were acquired in the range from 5 to 4428 $cm^{-1}$.

## 3. Results and discussion

$Al_2O_3$ insulator has been grown by thermal ALD directly on pristine 1L EG surface, without any type of pre-functionalization or seeding-layer. In order to monitor the morphological evolution of the deposited $Al_2O_3$ from the early stages of nucleation to the formation of a continuous film, a series of depositions has been carried out with increasing number of the ALD cycles in the range from 10 to 190 cycles. Fig.1 reports four representative AFM morphologies of the as-grown EG (a) and of the deposited $Al_2O_3$ after 10 (d), 40 (g) and 80 (l) ALD cycles.

The as-grown sample exhibits a very flat surface, with large terraces separated by sub-nanometer steps associated to the 4H-SiC substrate. The low surface roughness of EG residing on the SiC terraces is also demonstrated by the representative line-scan (b) and the very narrow distribution of the height values (c), fitted with a single Gaussian peak with full-width-half-maximum (FWHM) of 0.2 nm.

Fig.1(d) shows an incomplete surface coverage, with a large density of $Al_2O_3$ islands, in the case of the 10-cycles sample. A line-scan along the dashed red line of the AFM map is reported in Fig.1(e), from which a typical height of the $Al_2O_3$ islands in the order of 0.6 – 0.8 nm can be deduced. The fraction of EG surface covered by the insulator was estimated from the height distribution (Fig.1(f)) extracted from the morphology map. Such broad distribution was fitted with two Gaussian peaks, associated to the uncovered and covered EG regions, respectively. By integration of the counts under each of them, 14% and 86% percentages for the uncovered and covered areas were evaluated. Furthermore, an average height of the $Al_2O_3$ highlands of 0.7 nm was deduced from the separation of the two Gaussian peaks.

As expected for the early stages of ALD processes on a $sp^2$ graphene surface, island growth is observed rather than an ideal layer-by-layer 2D growth. However, in the present case of 1L EG/SiC, it is worth noting a large density and uniform distribution of the $Al_2O_3$ islands, originating from the growth of the initial nucleation sites. This scenario is very different from the one typically observed in the cases of exfoliated or CVD graphene transferred to a substrate, where the nucleation sites are mainly localized on the edges, on structural defects or on the wrinkle regions. The high and uniform density of nucleation sites in the case of 1L EG can be ascribed to the homogeneously distributed electrostatic forces generated by the buffer layer at EG/SiC interface [28].

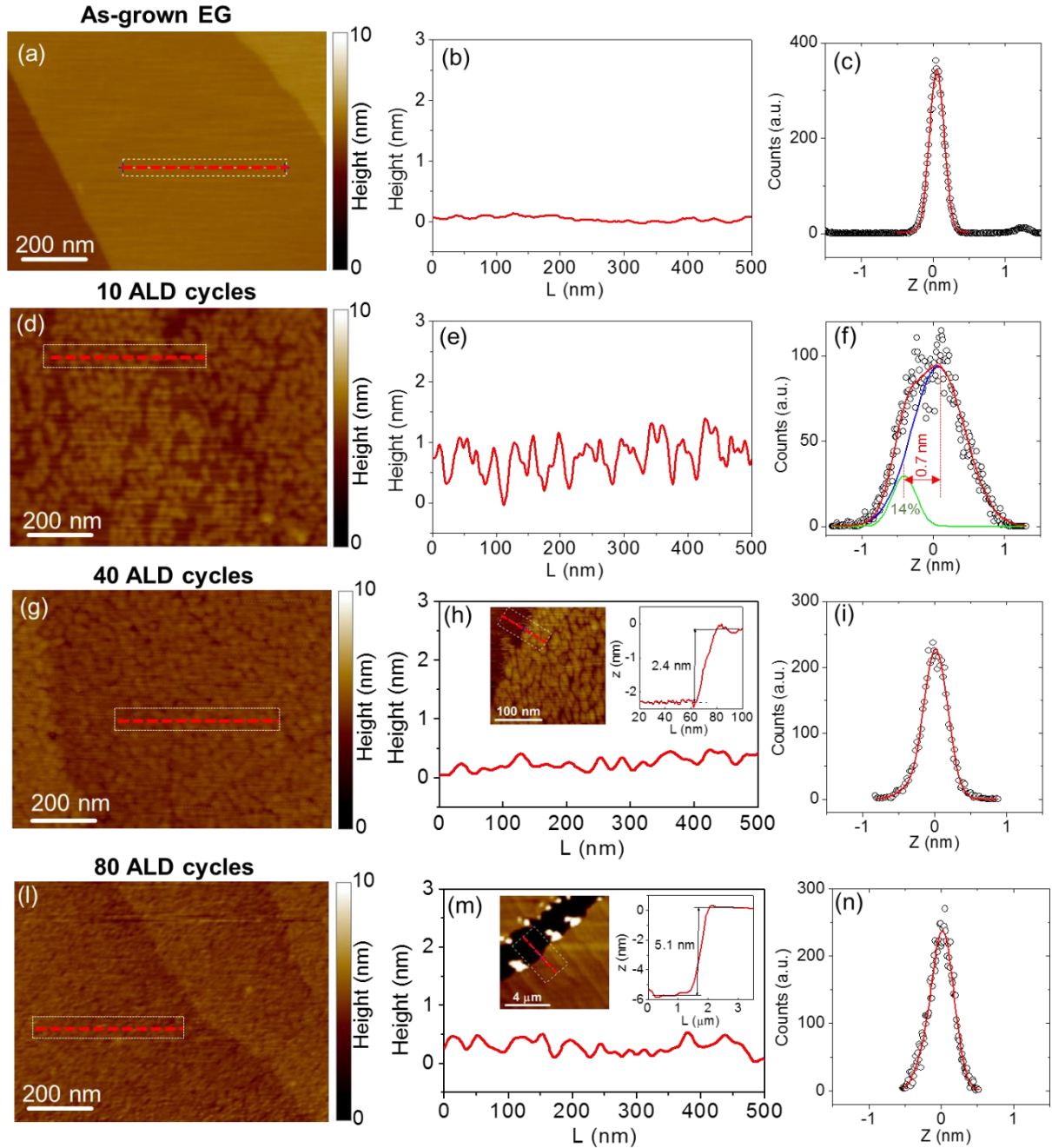

**Figure 1.** AFM morphology of the as-grown EG surface (a) and after 10 (d), 40 (g) and 80 (l) ALD cycles for the $Al_2O_3$ deposition. Representative height line-scans for the as-grown (b), 10-cycles (e), 40-cycles (h) and 80-cycles (m) sample. An incomplete coverage by $Al_2O_3$ islands with typical 0.6-0.8 nm height is observed after 10 ALD cycles. Closed $Al_2O_3$ layers are obtained after 40 and 80 ALD cycles, and the layer thickness was evaluated by the step height s measured on a scratch of the layers (inserts of panels (h) and (m)). Distributions of the height values extracted from the morphology maps for the as-grown (c), 10-cycles (f), 40-cycles (i) and 80-cycles (n) sample.

Fig.1(g) shows that, after only 40 ALD cycles, the full coverage of the EG surface was achieved due the coalescence of the $Al_2O_3$ islands. A much smoother morphology than for the 10-cycles sample can be deduced from the representative line profile in Fig.1(h) and from the narrower height distribution in Fig.1(i), fitted with a single Gaussian peak with FWHM of 0.33 nm. The deposited $Al_2O_3$ film thickness (~2.4 nm) was estimated in this case by measuring the step height of an intentionally scratched region, as illustrated in the insert of Fig. 1(h).

After 80 ALD cycles, the deposited $Al_2O_3$ insulator reaches the properties of a compact film perfectly conformal with the atomic steps of the substrate, as shown in the AFM image in Fig.1(l). The line profile in Fig.1(m) resembles the one previously measured on the 40-cycles sample, but with a slightly smoother morphology, as deduced by the smaller FWHM=0.28 nm of the single Gaussian height distribution in Fig.1(n). This suggests the occurrence of conformal layer-by-layer growth on the $Al_2O_3$ film formed by island coalescence. Finally, a film thickness of ~5.1 nm was evaluated from the step height on a scratched region, as shown in the insert of Fig.1(m).

Fig.2(a) summarizes the percentage of the covered EG surface by $Al_2O_3$ as a function of the number of ALD cycles. The covering level was beyond 80% already after 10 deposition cycles and full coverage (100%) was achieved after 40 deposition cycles, by coalescence of the growing $Al_2O_3$ islands. Fig.2(b), left scale, illustrates the increase of the $Al_2O_3$ island height with the number of ALD cycles in the regime of island growth (i.e. from 0 to 20 ALD cycles). Furthermore, Fig.2(b), right scale, shows the evolution of the film thickness with the number of cycles, after the formation of a continuous $Al_2O_3$ layer (i.e. above 40 ALD cycles). Although both the island height and the film thickness exhibit a linear increase with the number of cycles, different growth rates (evaluated as the slope of the fit) can be observed in the two regimes. This difference indicates a predominant 3D growth mechanism in the early stages of the deposition, followed by a 2D layer-by-layer growth after the formation of a continuous $Al_2O_3$ film (i.e. starting from 40 ALD cycles).

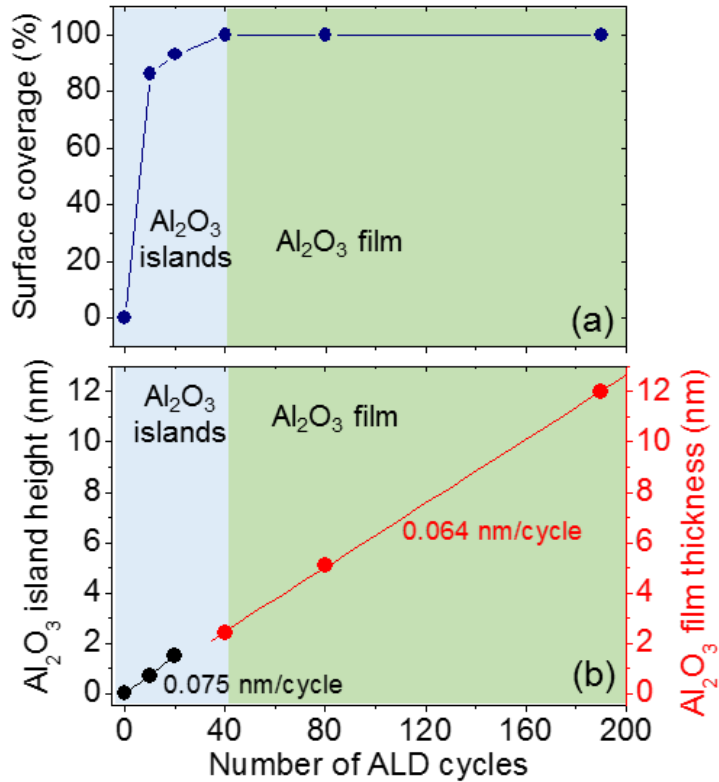

**Figure 2.** (a) Percentage of the EG surface covered by the deposited $Al_2O_3$ as a function of the number of ALD cycles. (b) Height of $Al_2O_3$ islands (left scale) and $Al_2O_3$ film thickness (right scale) versus the number of ALD cycles.

The island growth and coalescence phases have been further investigated by analyzing the evolution of the island lateral dimensions in the samples from 10 to 40 ALD cycles. Fig.3(a) and (c) shows two representative high resolution AFM maps for the 10 and 40 cycles samples, where the island contours have been properly highlighted. The distributions of island diameters evaluated by statistical analysis of these images are reported in Fig.3(b) and (d), respectively. The 10-cycles sample (Fig.3(b)) exhibits a bimodal distribution of diameters, that has been fitted by two Gaussian peaks corresponding to 7 nm and 20 nm average island sizes. A much broader islands diameter distribution has been found for the 40-cycles sample, as illustrated by the histogram in Fig.3(d). Interestingly, the multi-peaks Gaussian fit of this distribution includes both peaks at ~7 nm and ~20 nm (similarly to the 10-cycles sample), and further two peaks at ~30 nm and ~50 nm. This means that in the 40-cycles process there are islands of insulator further expanded on the EG substrate.

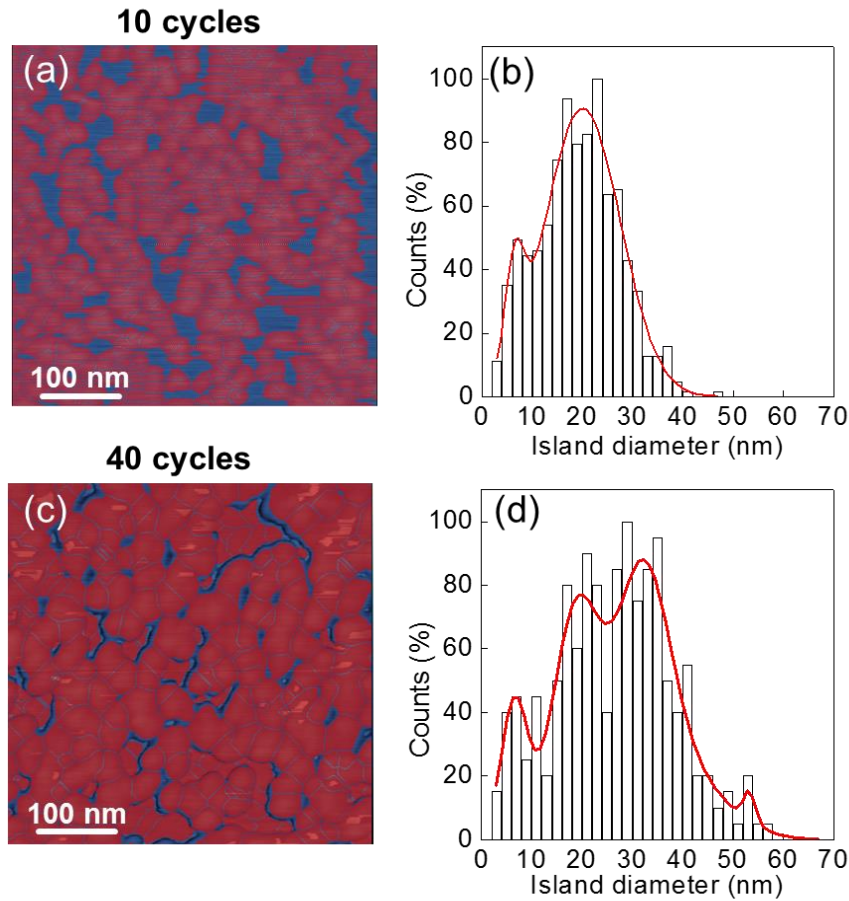

**Figure 3** AFM maps for the 10 ALD cycles (a) and 40 ALD cycles (c) samples, with the contours of the $Al_2O_3$ islands properly highlighted to perform statistical analysis of their lateral size. Distributions of the island diameters for the 10 ALD cycles (b) and 40 ALD cycles (d) samples.

Monitoring the island size distribution during the ALD growth provides an insight into the mechanisms of nucleation and coalescence, as demonstrated in recent experimental and modeling works [34,35]. Besides the deposition parameters (temperature, precursors), the shape of the island size distribution is strongly dependent on the surface energy of the substrate. In particular, in the case of hydrophobic substrates with low surface energy, diffusion and aggregation phenomena have been shown to play an important role in the island size evolution during ALD processes [36]. In the specific case of graphene, where physisorption is the main adsorption mechanism during ALD nucleation, the weak interactions between adsorbed material and graphene can be the source of surface diffusion phenomena. Hence, in the early stages of the ALD process, there is a parallel deposition of atoms on the bare graphene surface and on the pre-existing islands, combined with the diffusion of atoms or entire islands and dynamic coalescence phenomena [37,38].

Based on these considerations, the evolution of the morphology of the deposited $Al_2O_3$ on EG can be explained according to the schematic depicted in Fig.4.

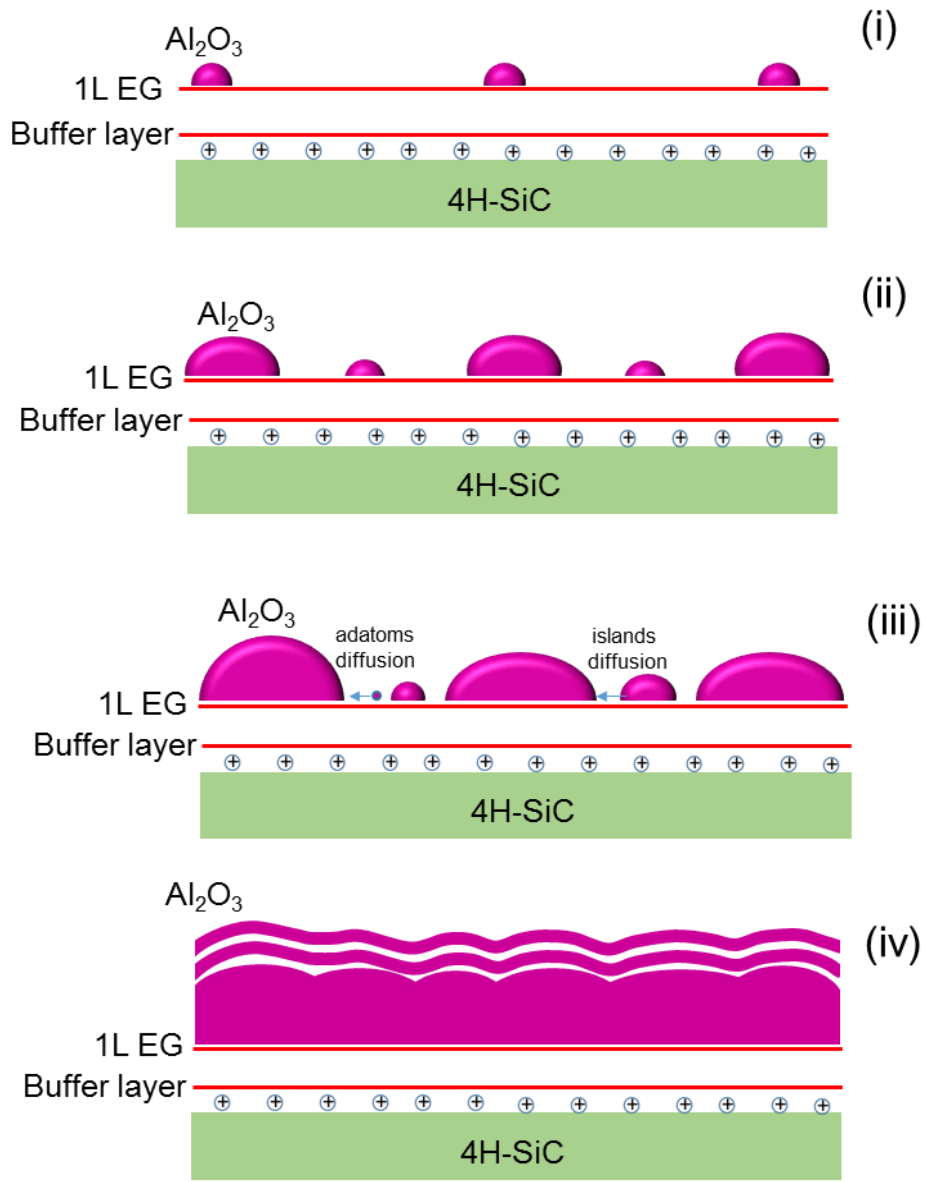

**Figure 4** Schematic representation of the evolution of the deposited $Al_2O_3$ morphology on EG with increasing number of ALD cycles

It displays four steps of the evolution with increasing the number of ALD cycles (from top down):

(i) the formation of the first nuclei on the Graphene surface;

(ii) the adsorption of atoms on the pre-existent islands (resulting in growth both in the lateral and vertical dimensions) and the nucleation of new islands on the bare Graphene regions;

(iii) the growth of larger islands at expenses of the smaller ones (Ostwald ripening) [15] mediated by adatoms diffusion on the graphene surface, or the dynamic coalescence due to the diffusion and merging of entire islands;

(iv) the layer-by-layer ALD growth, after the formation of a continuous film on graphene.

The impact of the thermal ALD process on the doping and strain of EG has been evaluated by micro-Raman spectroscopy. Fig.5 shows a set of representative Raman spectra collected on the SiC substrate, on the as-grown EG and after $Al_2O_3$ deposition using an increasing number of ALD cycles (from 10 to 190).

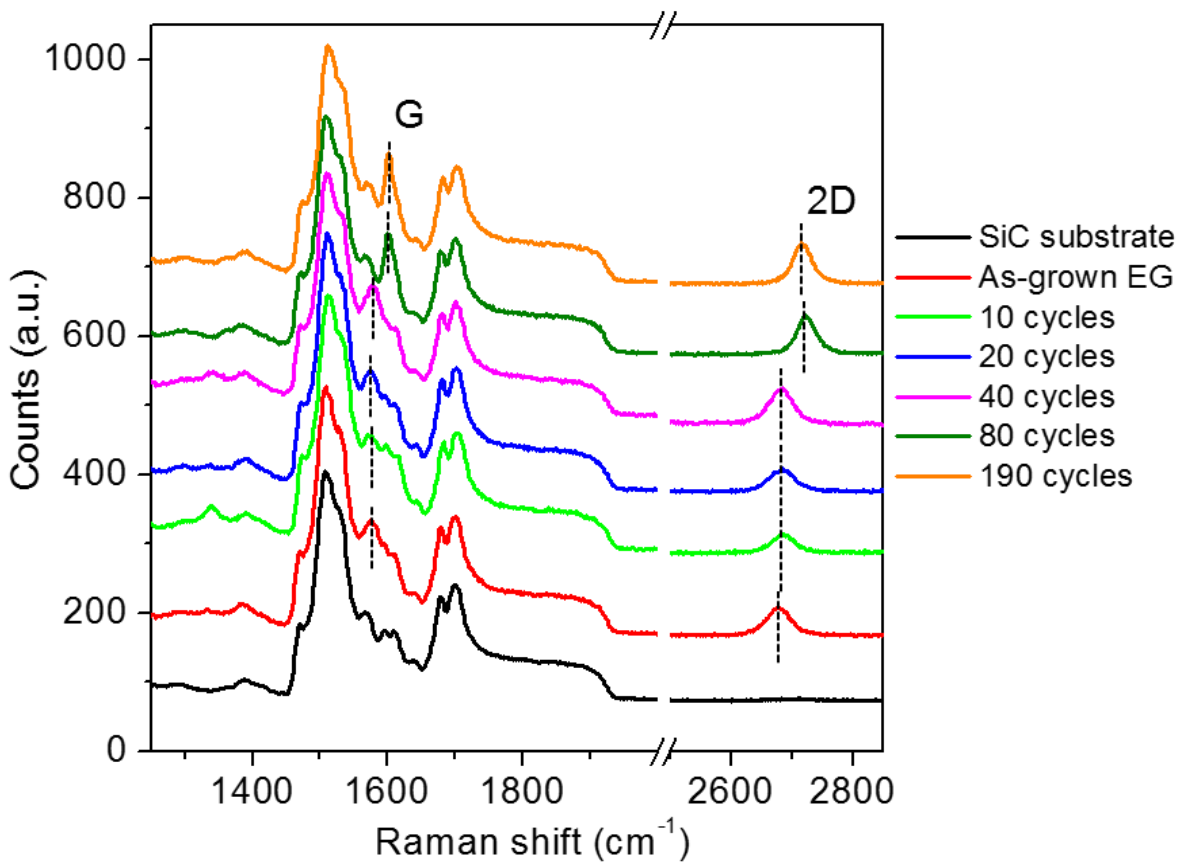

**Figure 5** Raman spectra collected on the SiC substrate, on the as-grown EG and after $Al_2O_3$ deposition with 10, 20, 40, 80 and 190 ALD cycles. The positions of the characteristic G and 2D peaks are indicated by vertical dashed lines.

All the spectra have been normalized to the intensity of the Raman peak of the SiC substrate. The characteristic 2D and G peaks associated with graphitic carbon can be observed, with the G peak partially overlapping the Raman features of the SiC substrate. Noteworthy, no significant changes can be observed between the as-grown and $Al_2O_3$ covered EG in the D peak spectral region (at ~1300 cm$^{-1}$), confirming that the ALD process does not introduce damages in EG. Furthermore, negligible shifts in the G and 2D peak positions occur moving from pristine EG up to 40 ALD cycles, whereas a significant blue shift of both peaks can be observed after 80 cycles. To better elucidate the origin of this shift, several Raman spectra have been collected at different positions on the as-grown EG samples and on all the samples with $Al_2O_3$ deposition at increasing number of cycles. Fig.6(a) shows a correlative plot of the 2D and G peak positions extracted from the Raman spectra on the different samples. The 2D and G positions for the ideal (i.e. strain-free and undoped) graphene ($\omega_{2D,0}$ = 2670 cm$^{-1}$; $\omega_{G,0}$ =1582 cm$^{-1}$) [39] have been also reported in the plot, along with the line describing the theoretical behavior of purely strained graphene [39] and the doping line evaluated from literature data for unstrained n-type doped graphene [40]. The point clouds for as-grown EG and for the samples with 10 to 40 ALD cycles exhibit a broad distribution around the strain line and almost overlapp each other. On the other hand, a narrower distribution and a shift along the strain line toward higher compressive strain values is observed after 80 and 190 ALD cycles.

The plot in Fig.6(a) indicates that the average doping of EG is not significantly affected by the $Al_2O_3$ ALD process, whereas an increase of the EG compressive strain is observed for a few ALD cycles higher than 80. Fig.6(b) reports the histograms of the strain distribution for all the considered samples, evaluated from the correlative plot using the geometrical analysis described in Refs. [39,41]. Very similar broad distributions with an average compressive strain $\varepsilon$=-0.15% can be observed up to 40 ALD cycles, whereas a narrower distribution with the peak at $\varepsilon$=-0.35% is found after 80 and 190 ALD cycles. The evolution of the compressive strain as a function of the number of ALD cycles is also summarized in the plot of Fig.6(c), where the points represent the average strain values and the error bars the standard deviations for each histogram in Fig.6(b).

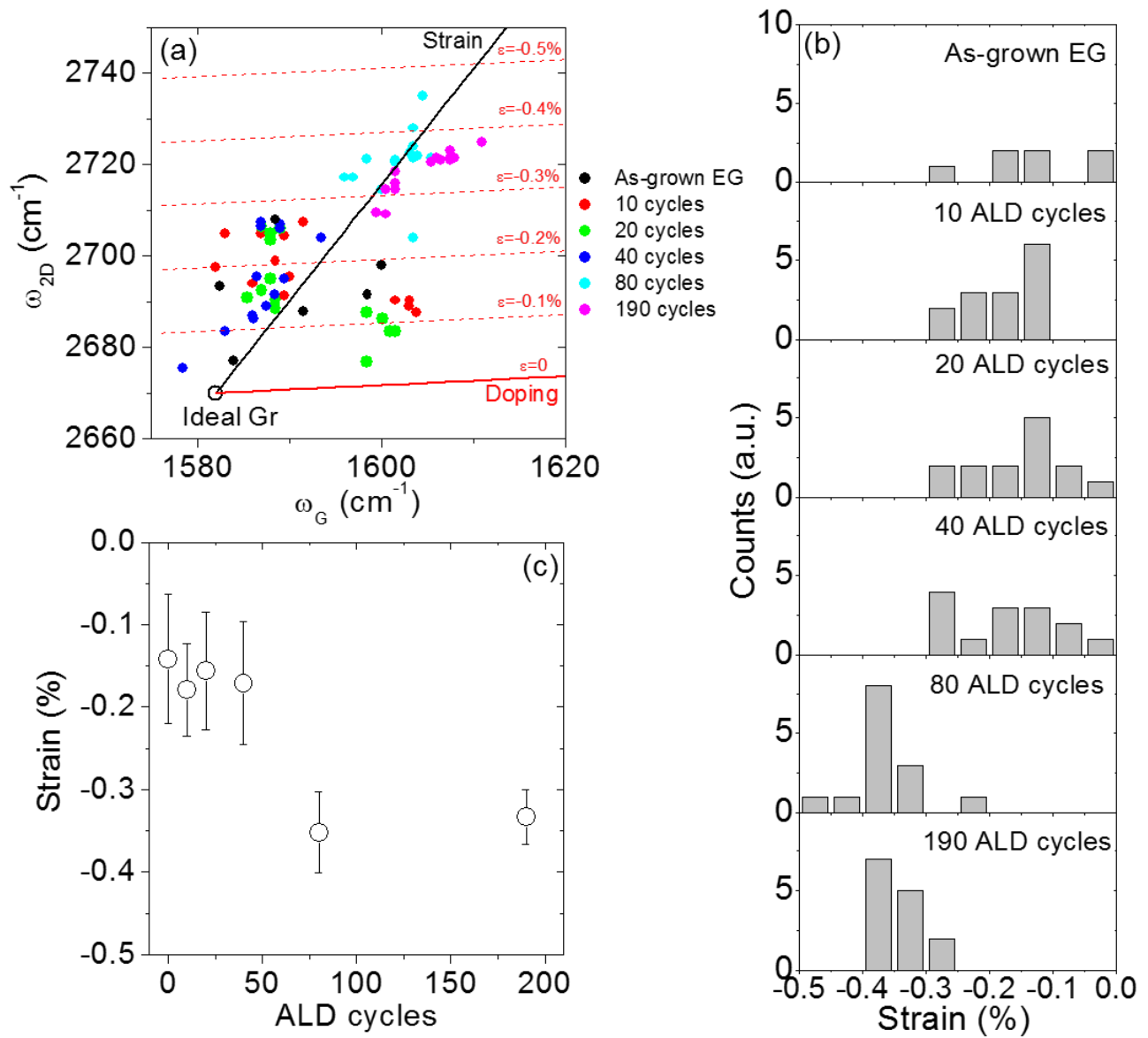

**Figure 6** (a) Correlative plot of the 2D vs G peak wavenumbers ($\omega_{2D}, \omega_G$) from Raman spectra collected at several positions on the as-grown EG sample and on the samples subjected to increasing number of $Al_2O_3$ ALD cycles. The ideal (strain-free and undoped) graphene is represented by the point ($\omega_{2D,0} = 2670$ cm$^{-1}$, $\omega_{G,0} = 1582$ cm$^{-1}$). The theoretical strain line (black) and the doping line for n-type graphene (red) are also reported in the plot. (b Histograms of the strain values for all the considered samples evaluated by the correlative plot. (c) Evolution of the compressive strain as a function of the number of ALD cycles.

Such a behavior, deduced from the analysis of the characteristic graphene Raman peaks, is consistent with the nanoscale resolution morphological analyses by AFM. In particular, the native compressive strain of EG appears to be almost unaffected by the early stages of the ALD process, i.e. during the nucleation and growth of $Al_2O_3$ islands, when the insulating film coverage is still incomplete. No significant strain variations are observed also after 40 ALD cycles, when the AFM analyses showed full coverage of EG with a ~2.4 nm $Al_2O_3$ film obtained from the coalescence of $Al_2O_3$ islands. This indicates that the insulating film was still not compact enough to influence the strain of the underlying EG. Finally, the factor of two increase of the compressive strain after 80 ALD cycles indicates that the $Al_2O_3$ film (with 5.1 nm thickness) reached a compact structure, resulting in deformation of the ultrathin graphene layer underneath. Noteworthy, after the insulator/EG structure is stabilized, no further increase in the compressive strain is observed for larger number of ALD cycles.

4. Conclusion

In conclusion, the ALD nucleation and growth mechanisms of ultra-thin $Al_2O_3$ films on monolayer EG on silicon carbide (4H-SiC) has been investigated in details by morphological analyses. Island growth from a large and uniform density of nucleation sites was observed in the early stages of the deposition, followed by the formation of a continuous $Al_2O_3$ film (~2.4 nm thick) due to the islands coalescence after only 40 ALD cycles. Afterwards, layer-by-layer growth giving rise to a conformal and compact $Al_2O_3$ film was observed. Raman spectroscopy analyses showed low impact of the ALD process on the defects density and on the average doping of EG. The compressive strain of as-grown EG ($\varepsilon$=-0.15%) was also almost unaffected during the nucleation, growth and coalescence of $Al_2O_3$ islands, whereas an increase of the strain by more than a factor of two ($\varepsilon$=-0.35%) was observed after the formation of a compact $Al_2O_3$ film. The obtained results can have important implications for device applications of epitaxial graphene.

**Acknowledgments**

The authors acknowledge S. Di Franco from CNR-IMM, Catania, for his expert technical assistance with samples preparation. This project has been supported, in part, by MIUR in the framework of the FlagERA-JTC 2019 project ETMOS. RY is acknowledging the support by the SSF project GMT14-0077 and VR 2018-04962.